\documentstyle[preprint,aps, epsfig]{revtex}

\begin{document}

\draft

\preprint{
\hfill$\vcenter{\hbox{\bf FERMILAB Pub-97/305-T}
                \hbox{\bf IFUSP-P 1271} 
                \hbox{\bf hep-ph/9709319}
             }$ }

\title{Signal and Backgrounds for Leptoquarks at the LHC}

\author{O.\ J.\ P.\ \'Eboli\thanks{Email: eboli@fma.if.usp.br}, 
R.\ Z.\ Funchal\thanks{Email: zukanov@charme.if.usp.br}, }

\address{Instituto de F\'{\i}sica, Universidade de S\~ao Paulo \\
C.P.\ 66.318, 05315-970 S\~ao Paulo, Brazil}

\author{T.\ L.\ Lungov\thanks{Email: lungov@fnal.gov} }

\address{Fermi National Accelerator Laboratory\\
P.O.\ Box 500, Batavia, IL 60510, USA}

\date{August 29, 1997}

\maketitle

\begin{abstract}
  
  We study the potentiality of the CERN Large Hadron Collider (LHC) to
  unravel the existence of first generation scalar leptoquarks.
  Working with the most general $SU(2)_L \otimes U(1)_Y$ invariant
  leptoquark interactions, we analyze in detail the signals and
  backgrounds that lead to a final state containing a pair $e^+e^-$
  and jets.  Our results indicate that a machine like the LHC will be
  able to discover leptoquarks with masses up to 2--3 TeV depending on
  their couplings.

\end{abstract}


\newpage


\section{Introduction}

Many extensions of the Standard Model (SM) treat quarks and leptons in
the same basis and, consequently, allow the existence of particles,
called leptoquarks, that mediate quark-lepton transitions.  The class
of theories exhibiting these particles includes composite models
\cite{comp,af}, grand unified theories \cite{gut}, technicolor models
\cite{tec}, and superstring-inspired models \cite{e6}.  Leptoquarks
carry simultaneously leptonic and baryonic number, and can have spin
$0$ or $1$.

Since leptoquarks are an undeniable signal of physics beyond the SM,
there have been several direct searches for them in accelerators.  So
far, all searches led to a negative result, with the possible
exception of some intriguing events in $e^+p$ collisions at HERA
\cite{heranew}. Analyzing the decay of the $Z$ into a pair of on-shell
leptoquarks, the LEP experiments established a lower bound $ M_{lq}
\gtrsim 44$ GeV for scalar leptoquarks \cite{LEP,DELPHI}. This limit
can be improved by looking for $Z$ decays into an on-shell leptoquark
and an off-shell one, resulting in $M_{lq} \gtrsim 65-73$ GeV
\cite{DELPHI}.  The search for scalar leptoquarks decaying exclusively
into electron-jet pairs at the Tevatron constrained their masses to be
$M_{lq} \gtrsim 225$ GeV \cite{PP}. Furthermore, the experiments at
HERA \cite{HERA} placed limits on their masses and couplings,
establishing that $M_{lq} \gtrsim 216-275$ GeV depending on the
leptoquark type and couplings.

The direct search for leptoquarks with masses above a few hundred GeV
can be carried out only in the next generation of colliders. There
have been many studies of the production of leptoquarks in the future
$pp$ \cite{fut:pp}, $ep$ \cite{buch,fut:ep}, $e^+e^-$ \cite{fut:ee},
$e^-e^-$ \cite{fut:elel}, $e\gamma$ \cite{fut:eg}, and $\gamma\gamma$
\cite{fut:gg} colliders.  In this work, we study the capability of the
CERN Large Hadron Collider (LHC) to unravel the existence of scalar
leptoquarks through the final state topology two jets plus a pair
$e^+e^-$. This is accomplished by a careful analyses of the signal and
its respective backgrounds. We performed our analyses for first
generation leptoquarks that couple to pairs $e^- u$, $e^+ u$, $e^- d$,
or $e^+ d$ with the leptoquark interactions described by the most
general effective Lagrangian that is invariant under $SU(3)_C \otimes
SU(2)_L \otimes U(1)_Y$ \cite{buch}.

It is interesting to notice that pair production of scalar leptoquarks
in a hadronic collider is essentially model independent since the
leptoquark--gluon interaction is fixed by the $SU(3)_C$ gauge
invariance. On the other hand, single production is model dependent
because it takes place via unknown interactions.  Notwithstanding,
these two signals are complementary because they allow us not only to
reveal their existence but also to determine their properties such as
mass and Yukawa couplings to quarks and leptons.

The single and pair productions of leptoquarks can give rise to jets
accompanied by $e^+e^-$ pairs. In this work, we performed a careful
analyses of all possible QCD and electroweak backgrounds for the jets
plus a $e^+e^-$ pair topology using the event generator PYTHIA version
5.7 \cite{pyt}, taking into account initial and final state QCD
radiation. The signal was also generated using this package.  We
devised a series of cuts not only to reduce the background and enhance
the signal but also to separate the single and pair production
mechanisms.  Our analyses improves the previous ones \cite{atlas}
since we considered all possible backgrounds as well as the most
general model for leptoquarks.

We analyzed the single leptoquark production taking into account the
contribution from pair production, which enhances the signal, and we
obtained exclusion regions in terms of the leptoquark mass and Yukawa
coupling. With a luminosity of 10 ($100$) fb$^{-1}$ and a
center-of-mass energy of $14$~TeV, the single leptoquark search at the
LHC can exclude their existence at 95\% CL for masses smaller than
$1.3$--$1.5$ (1.7--2.0) TeV irrespective of their Yukawa couplings.
For larger masses, the LHC reach depends on the coupling of the
leptoquark to fermions. On the other hand, the production of a
leptoquark pair can be observed up to masses of 1.1--1.3 (1.5--1.7)
TeV depending only upon the leptoquark type.

Low-energy experiments lead to strong indirect bounds on the couplings
and masses of leptoquarks, which can be used to define the goals of
new machines to search for these particles.  The main sources of
indirect constraints are:

$\bullet$ Leptoquarks give rise to Flavor Changing Neutral Current
(FCNC) processes if they couple to more than one family of quarks or
leptons \cite{shanker,fcnc}. In order to avoid strong bounds from
FCNC, we assumed that the leptoquarks couple to a single generation of
quarks and a single one of leptons. However, due to mixing effects on
the quark sector, there is still some amount of FCNC left
\cite{leurer} and, therefore, leptoquarks that couple to the first two
generations of quarks must comply with some low-energy bounds
\cite{leurer}.

$\bullet$ The analyses of the decays of pseudoscalar mesons, like the
pions, put stringent bounds on leptoquarks unless their coupling is
chiral -- that is, it is either left-handed or right-handed
\cite{shanker}.

$\bullet$ Leptoquarks that couple to the first family of quarks and
leptons are strongly constrained by atomic parity violation
\cite{apv}.  In this case, there is no choice of couplings that avoids
the strong limits.

$\bullet$ The analyses of the effects of leptoquarks on the $Z$
physics through radiative corrections lead to limits on the masses and
couplings of leptoquarks that couple to top quarks \cite{gb,jkm}.

As a rule of a thumb, the low-energy data constrain the masses of
leptoquarks to be larger than $0.5$---$1$ TeV when their Yukawa
coupling is equal to the electromagnetic coupling $e$
\cite{leurer,jkm,davi}. Therefore, our results indicate that the LHC
can confirm these indirect limits and also expand them considerably.

The outline of this paper is as follows. In Sec.\ \ref{l:eff} we
introduce the $SU(2)_L \otimes U(1)_Y$ invariant effective Lagrangians
that we analyzed and its connection to the production cross sections
used in PYTHIA. A detailed discussion of all possible backgrounds is
exhibited in Sec.\ \ref{bckg}, while Sec.\ \ref{sign} contains the
main features of the signals for single and double leptoquark
productions. Our analyses are shown in Sec.\ \ref{resu} and we draw
our conclusions in Sec.\ \ref{conc}.


\section{Models for scalar leptoquark interactions}
\label{l:eff}

A natural hypothesis for theories beyond the SM is that they exhibit
the gauge symmetry $SU(2)_L \otimes U(1)_Y$ above the electroweak
symmetry breaking scale $v$, therefore, we imposed this symmetry on
the leptoquark interactions.  In order to avoid strong bounds coming
from the proton lifetime experiments, we required baryon ($B$) and
lepton ($L$) number conservation, which forbids the leptoquarks to
couple to diquarks.  The most general effective Lagrangian for
leptoquarks satisfying the above requirements and electric charge and
color conservation is given by \cite{buch}
\begin{eqnarray}
{\cal L}_{\text{eff}}~  &  & =~ {\cal L}_{F=2} ~+~ {\cal L}_{F=0} 
\; , 
\label{e:int}
\\
{\cal L}_{F=2}~  &  & =~ g_{\text{1L}}~ \bar{q}^c_L~ i \tau_2~ 
\ell_L ~S_{1L}+ 
g_{\text{1R}}~ \bar{u}^c_R~ e_R ~ S_{1R} 
+ \tilde{g}_{\text{1R}}~ \bar{d}^c_R ~ e_R ~ \tilde{S}_1
+ g_{3L}~ \bar{q}^c_L~ i \tau_2~\vec{\tau}~ \ell_L \cdot \vec{S}_3 
\; ,
\label{lag:fer}\\
{\cal L}_{F=0}~  &  & =~ h_{\text{2L}}~ R_{2L}^T~ \bar{u}_R~ i \tau_2 ~
 \ell_L 
+ h_{\text{2R}}~ \bar{q}_L  ~ e_R ~  R_{2R} 
+ \tilde{h}_{\text{2L}}~ \tilde{R}^T_2~ \bar{d}_R~ i \tau_2~ \ell_L
\; ,
\label{eff} 
\end{eqnarray}
where $F=3B+L$, $q$ ($\ell$) stands for the left-handed quark (lepton)
doublet, and $u_R$, $d_R$, and $e_R$ are the singlet components of the
fermions. We denote the charge conjugated fermion fields by
$\psi^c=C\bar\psi^T$ and we omitted in Eqs.\ (\ref{lag:fer}) and
(\ref{eff}) the flavor indices of the leptoquark couplings to
fermions. The leptoquarks $S_{1R(L)}$ and $\tilde{S}_1$ are singlets
under $SU(2)_L$, while $R_{2R(L)}$ and $\tilde{R}_2$ are doublets, and
$S_3$ is a triplet.  The quantum numbers for all scalar leptoquarks
can be found, for instance, in the last reference of \cite{fut:ee}.
Since leptoquarks are color triplets, it is natural to assume that
they interact with gluons.  To obtain their couplings to gluons we
substituted $\partial_\mu$ by the covariant derivative $D_\mu =
\partial_\mu + i g_s \lambda^a_{jk} G^a_\mu /2$ in the leptoquark
kinetic Lagrangian.

From the above interactions we can see that the main decay modes of
leptoquarks are into pairs $e^\pm q$ and $\nu_e q^\prime$, thus, their
signal is a $e^\pm$ plus a jet, or a jet plus missing energy. In this
work we considered only the first decay and took into account properly
the corresponding branching ratio. Moreover, the event generator
PYTHIA assumes that the leptoquark interaction with quarks and leptons
is described by
\begin{equation}
\bar{e} (a+b\gamma_5) q  \;,
\end{equation}
and the cross sections are expressed in terms of the parameter
$\kappa$ defined as
\begin{equation}
\kappa \alpha_{\text{em}}\equiv \frac{a^2 + b^2}{4\pi}
\label{eq:kap}
\end{equation}
with $\alpha_{\text{em}}$ being the fine structure constant.  We
exhibit in Table \ref{t:cor} the leptoquarks that can be analyzed
using the final state $e^\pm$ plus a jet, as well as, their decay
products, branching ratio into $e^\pm q$, and the correspondent value
of $\kappa$. As we can see from Eqs.\ (\ref{lag:fer}) and (\ref{eff}),
only the leptoquarks $R^2_{2L}$, $\tilde{R}^2_2$ and $S_3^-$ decay
exclusively into a jet and a neutrino.


\section{Backgrounds}
\label{bckg}

At the parton level, the single production of leptoquarks leads to a
final state exhibiting a pair $e^+e^-$ and $q$ ($\bar{q}$).  After the
parton shower and hadronization the final state usually contains more
than one jet, and consequently, the backgrounds for single and pair
productions of leptoquarks are basically the same.  In our analyses we
kept track of the $e^\pm$ (jet) carrying the largest transverse
momentum, that we denoted by $e_1$ ($j_1$), and the $e^\pm$ (jet) with
the second largest $p_T$, that we called $e_2$ ($j_2$).  The
reconstruction of the jets in the final state was done using the
subroutine LUCELL of PYTHIA, requiring the transverse energy of the
jet to be larger than 7 GeV inside a cone $\Delta R = \sqrt{ \Delta
  \eta^2 + \Delta \phi^2} =0.7$. Our calculations were performed using
the parton distribution functions of CTEQ2L \cite{cteq2l}.

Within the scope of the SM, there are many sources of backgrounds
leading to jets accompanied by a pair $e^+e^-$. We divided them into
three classes: QCD and electroweak processes and top quark production.

\subsection{QCD processes}

The reactions, included in the QCD class, depend exclusively on the
strong interaction and are listed in Table \ref{t:qcd1}, where $q$
($\bar{q}$) represents a (anti)quark of flavor $u$, $d$, $c$, $s$, or
$b$ and $g$ stands for a gluon. We also show in this table the total
cross section ($\sigma_{\text{bare}}$) for each process with the only
requirement that the partons produced in the hard scattering have a
transverse momentum larger than 100 GeV \cite{ckin}. The last column
in this table contains the total cross section
($\sigma_{\text{pair}}$) after imposing that the final state contains
a pair $e^+e^-$ with each lepton having a transverse momentum larger
than 50 GeV.

\begin{figure}
\begin{center}
\mbox{\epsfig{file=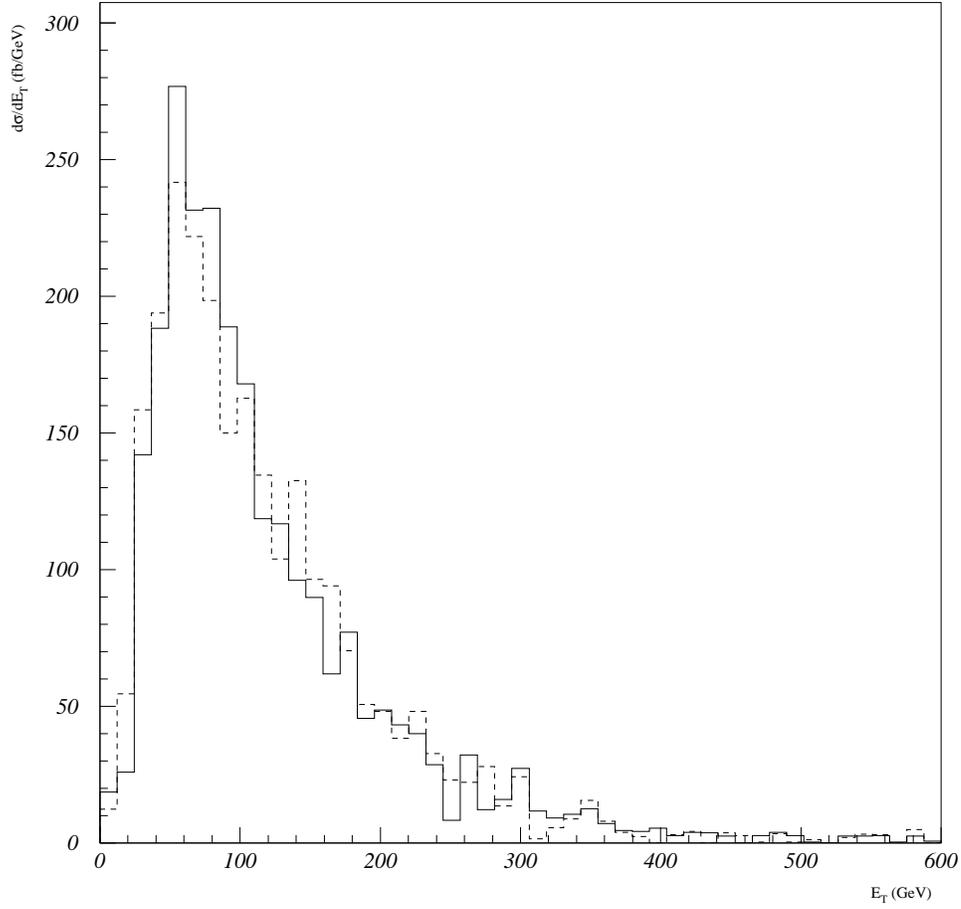,width=.85\linewidth}}
\end{center}
\caption{
  Hadronic transverse energy deposited in a cone of size $\Delta R =
  0.7$ around the direction of the $e_1$ (solid line) and the $e_2$
  (dashed line) in QCD events.}
\label{fig:qcd0}
\end{figure}

In this class of processes, the main source of hard $e^\pm$ is the
semileptonic decay of hadrons possessing quarks $c$ or $b$, which are
produced in the hard scattering or in the parton shower through the
splitting $g \rightarrow c \bar{c}$ ($b \bar{b}$). Since only a small
fraction of the decays give rise to hard leptons, we needed to improve
the efficiency of the Monte Carlo by decaying the same hadron several
times and keeping track of the fraction of events that yield the
desired signal. An important feature of the events in this class is
that close to the hard $e^\pm$ there is a substantial amount of
hadronic activity, as we can see from Fig.\ \ref{fig:qcd0}, where we
show the hadronic transverse energy deposited in a cone of size $R =
\sqrt{\Delta \eta^2 + \Delta \phi^2} = 0.7$ around the $e^\pm$
direction.

\begin{figure}
\begin{center}
\mbox{\epsfig{file=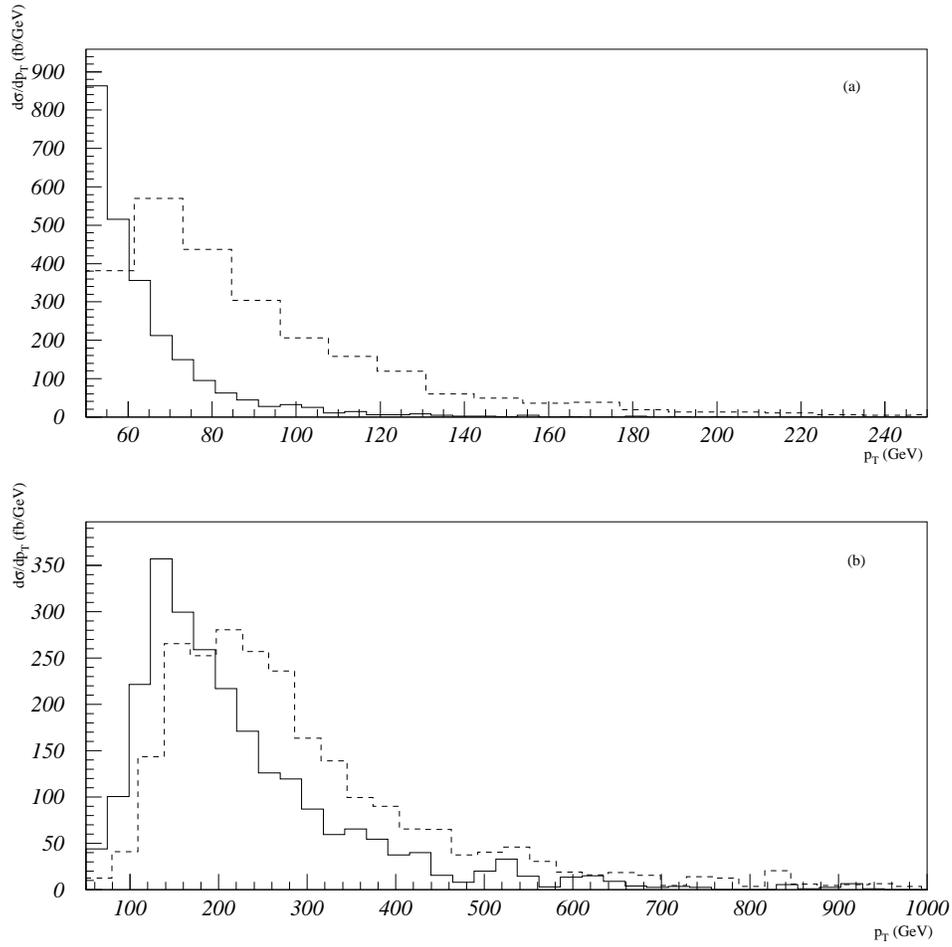,width=.85\linewidth}}
\end{center}
\caption{
  For QCD events: (a) the dashed (solid) line stands for the $p_T$
  distribution of $e_1$ ($e_2$); (b) the dashed (solid) line stands
  for the $p_T$ distribution of $j_1$ ($j_2$). }
\label{fig:qcd1}
\end{figure}

\begin{figure}
\begin{center}
\mbox{\epsfig{file=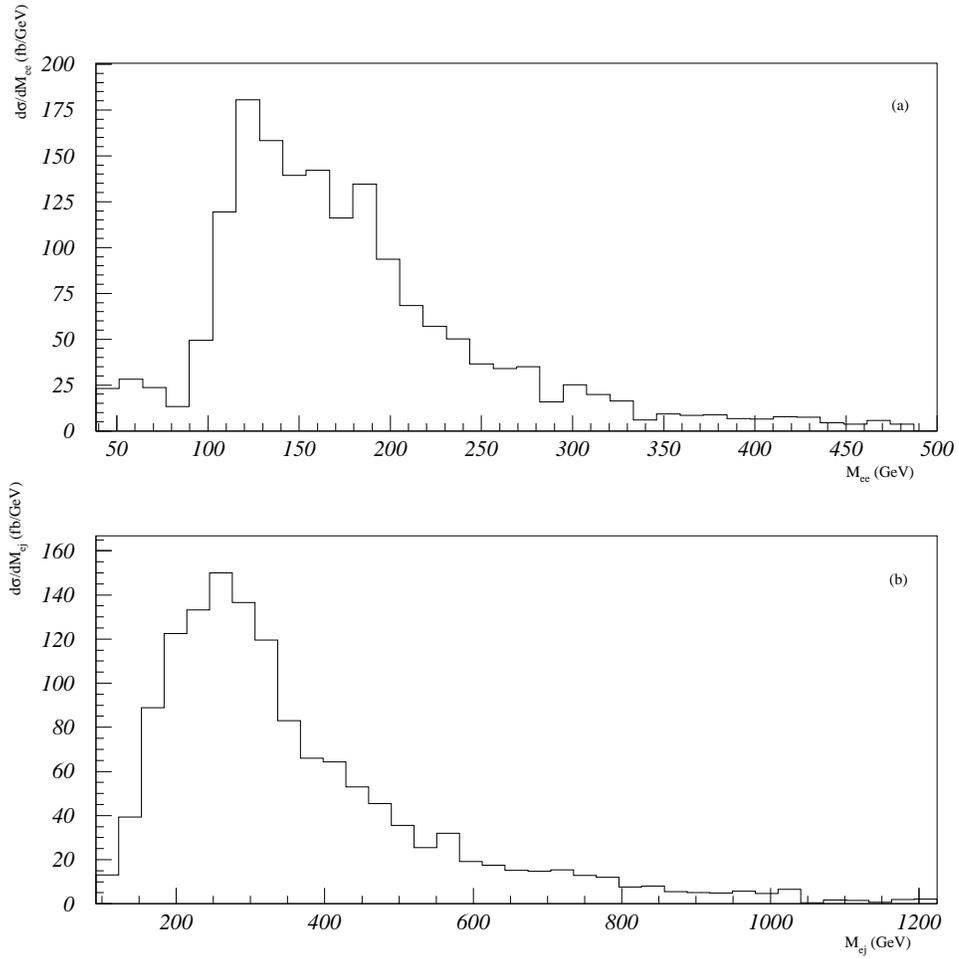,width=.85\linewidth}}
\end{center}
\caption{ 
  For QCD events: (a) $e_1 e_2$ invariant mass distribution; (b)
  $e^\pm$-jet invariant mass spectrum adding the 4 possible
  combinations.  }
\label{fig:qcd2}
\end{figure}

We show in Fig.\ \ref{fig:qcd1} the $p_T$ spectrum of the $e^\pm$ and
jets with the largest transverse momenta. As we can see from this
figure, the processes in this class yield a large number of jets with
large transverse momentum, however, the $p_T$ of the second most
energetic $e^\pm$ ($e_2$) is seldom larger than 200 GeV. From Fig.\ 
\ref{fig:qcd2}a we can learn the $e_1e_2$ invariant mass distribution
is peaked at small masses, however it still has a sizable value up to
invariant masses of the order of 400 GeV. We can assess the importance
of this background from the invariant mass spectrum of lepton-jet
pairs that is shown in Fig.\ \ref{fig:qcd2}b, where we added the
contributions from all $e_i$--$j_k$ pairs ($i,k = 1,2$).


\subsection{Electroweak processes}

\begin{figure}
\begin{center}
\mbox{\epsfig{file=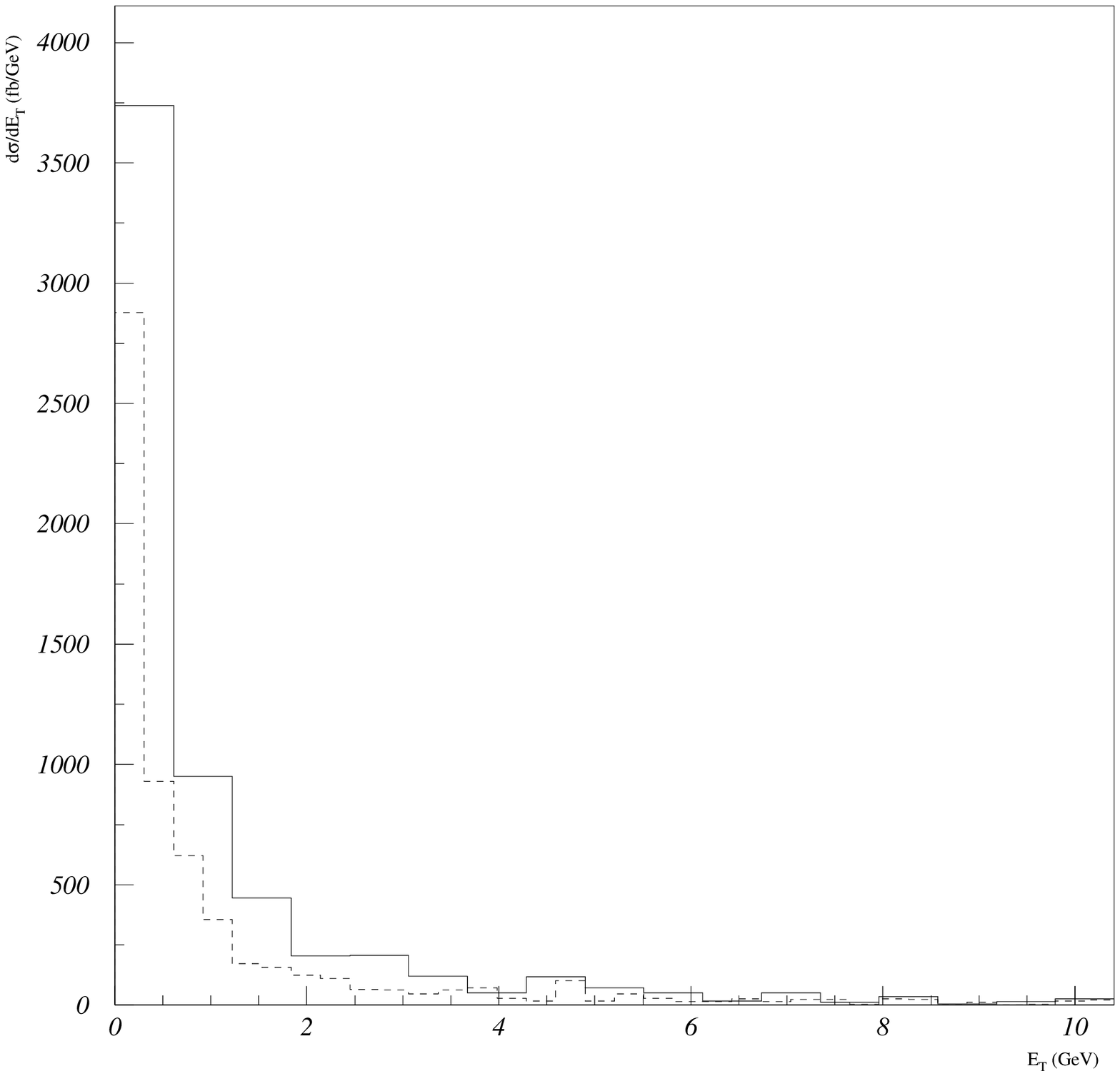,width=.85\linewidth}}
\end{center}
\caption{
  The same distributions of Fig.\ \protect\ref{fig:qcd0} for electroweak
  events.  }
\label{fig:ew0}
\end{figure}

The electroweak class contains the Drell-Yan production of quark pairs
and the single and pair productions of electroweak gauge bosons. We
list in Table \ref{t:ew1} all processes included into this class and
respective cross sections ($\sigma_{\text{bare}}$ and
$\sigma_{\text{pair}}$) evaluated with the same cuts applied to the
QCD processes. It is interesting to notice that the reaction $q_i g
\rightarrow q_i Z$ is the dominant background with the minimal cuts
applied so far. Moreover, the majority of the $e^\pm$ produced by this
class of processes originates from the decay of an electroweak gauge
boson, and consequently they tend to be isolated, as we can see from
Fig.\ \ref{fig:ew0}.

\begin{figure}
\begin{center}
\mbox{\epsfig{file=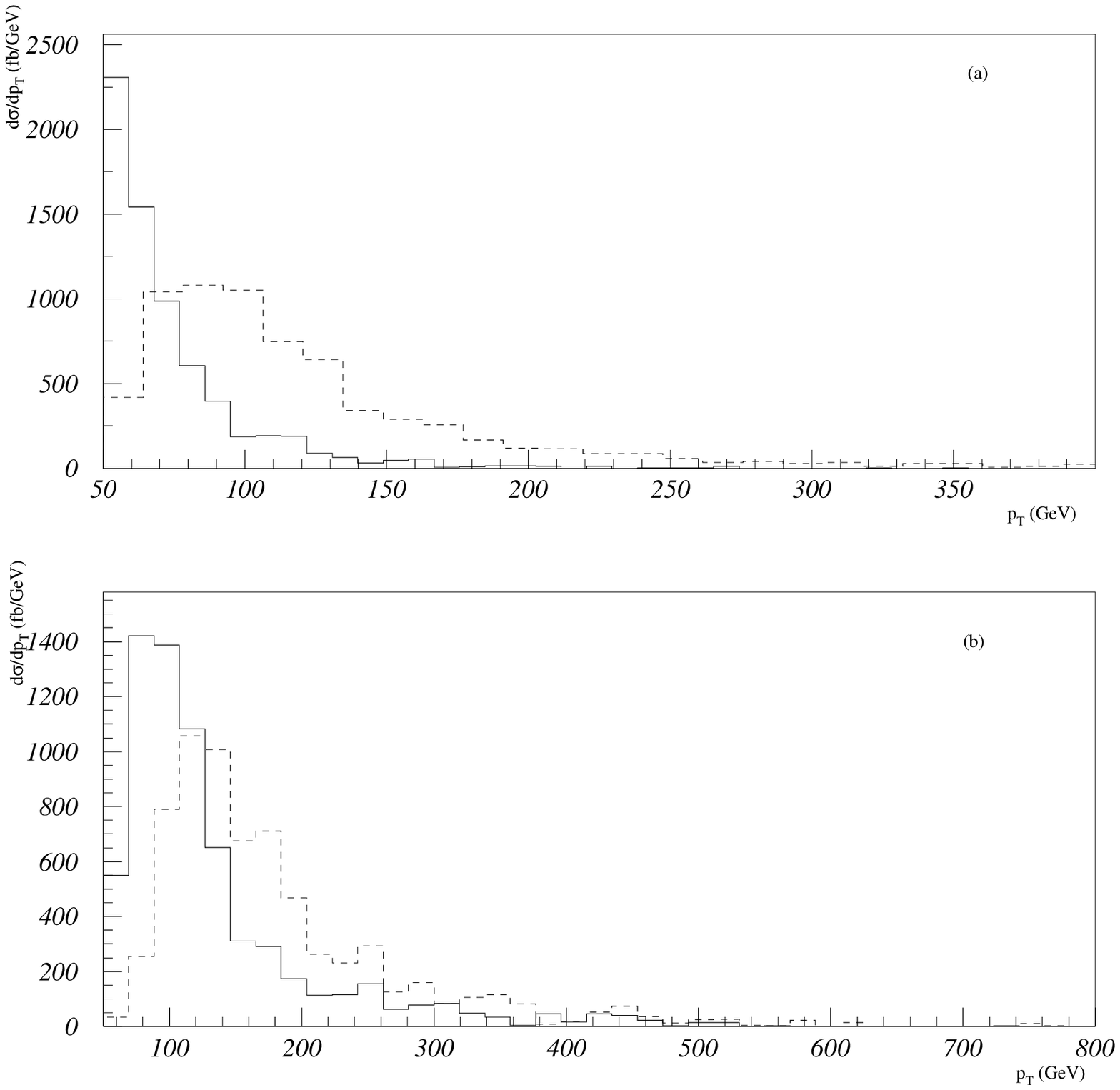,width=.85\linewidth}}
\end{center}
\caption{
  The same distributions of Fig.\ \protect\ref{fig:qcd1} for
  electroweak events.  }
\label{fig:ew1}
\end{figure}

\begin{figure}
\begin{center}
\mbox{\epsfig{file=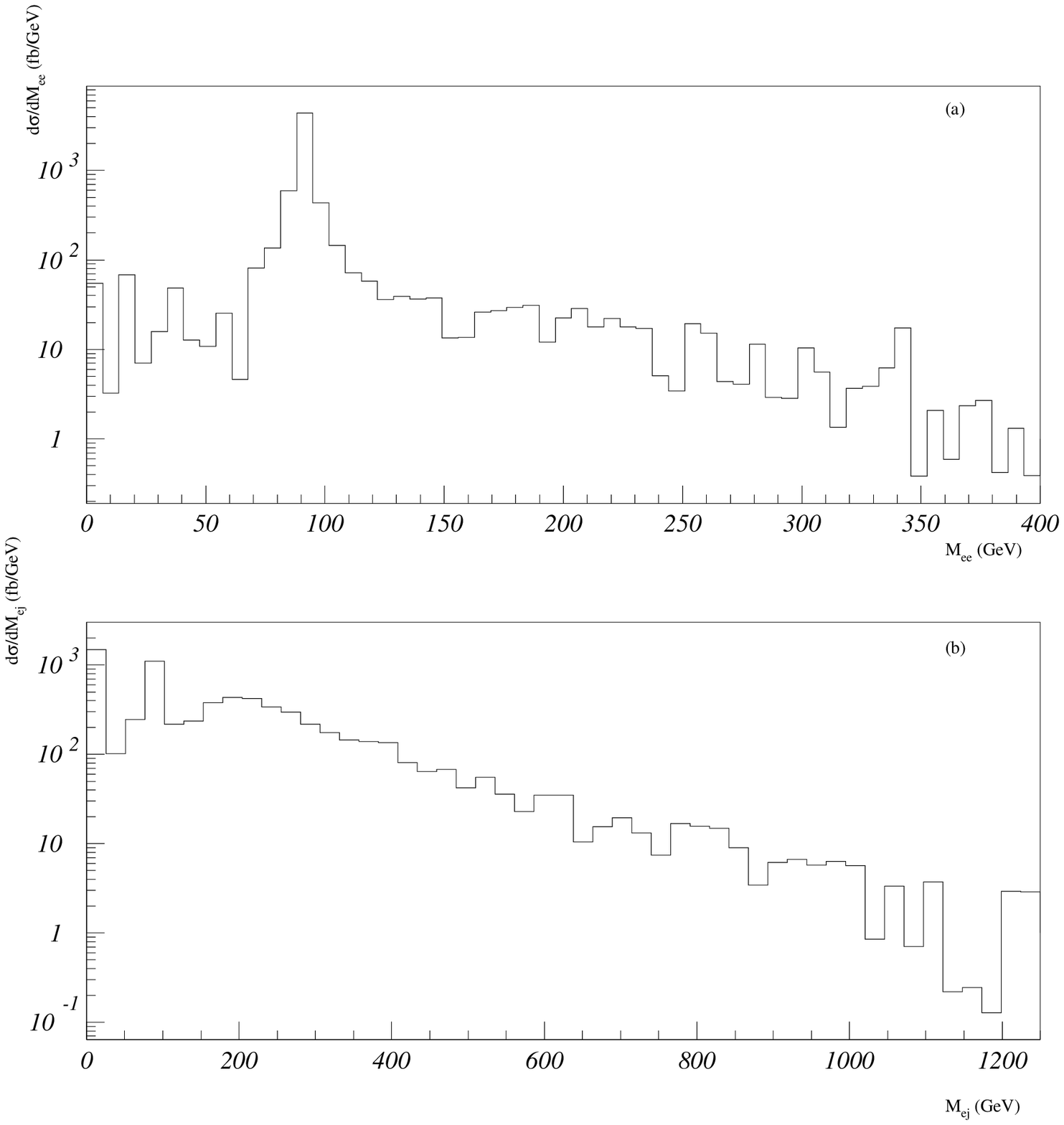,width=.85\linewidth}}
\end{center}
\caption{
  The same distribution of Fig.\ \protect\ref{fig:qcd2} for
  electroweak events.  }
\label{fig:ew2}
\end{figure}

We exhibit in Fig.\ \ref{fig:ew1} the $p_T$ spectrum of the $e^\pm$
and jets, which shows that this class of events yields leptons with
larger transverse momentum than the QCD class, while the jets have
smaller $p_T$. From Fig.\ \ref{fig:ew2}a we can learn that the $e_1
e_2$ invariant mass distribution is peaked around the $Z$ mass and its
is non-negligible up to invariant masses of the order of 400 GeV. We
plot in Fig.\ \ref{fig:ew2}b the sum of all $e_i$--$j_k$ invariant
masses, which indicates that the importance of these background
processes decrease for heavier leptoquarks.


\subsection{Top production}

\begin{figure}
\begin{center}
\mbox{\epsfig{file=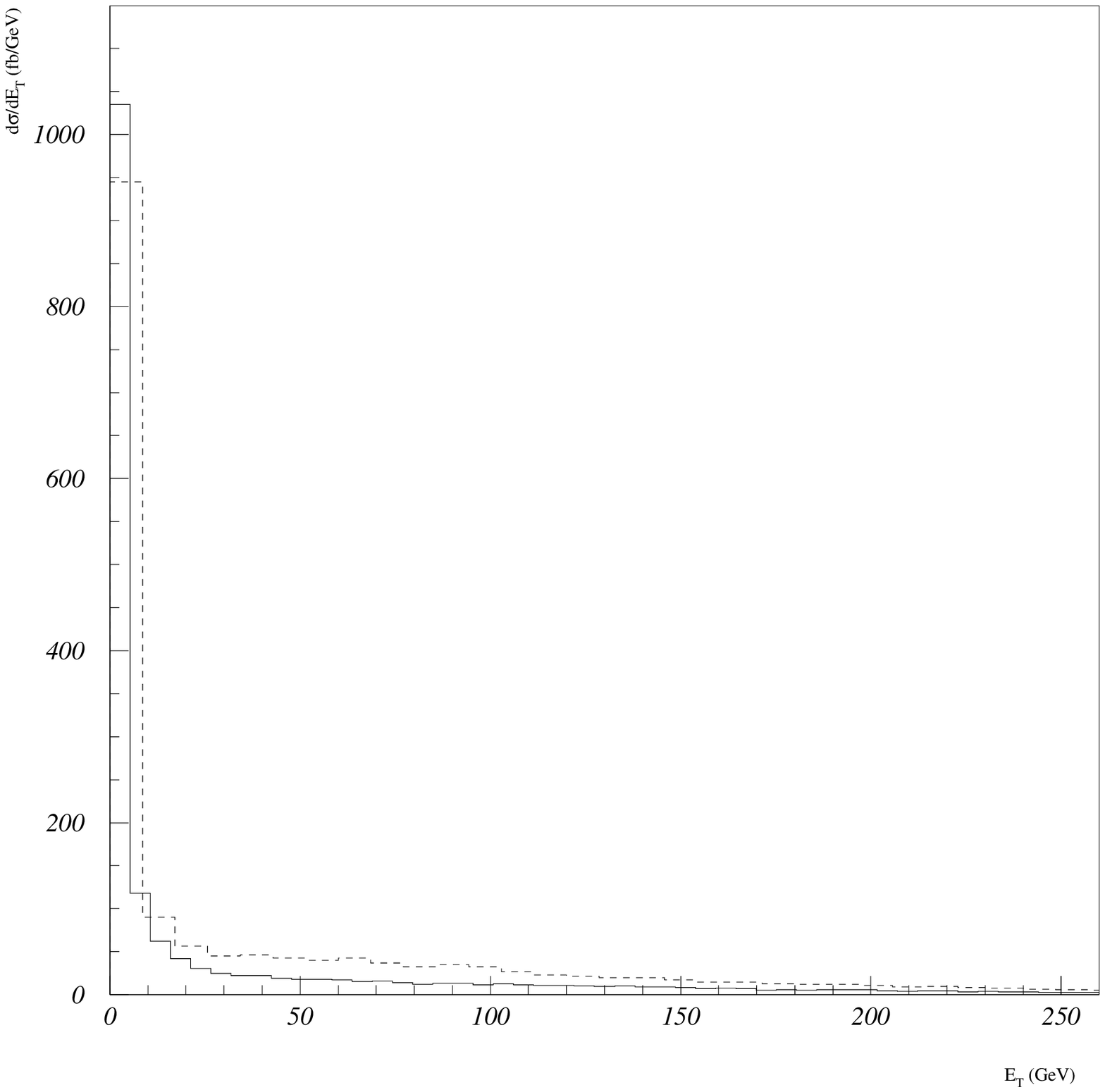,width=.85\linewidth}}
\end{center}
\caption{
  The same distribution of Fig.\ \protect\ref{fig:qcd0} for top
  production.  }
\label{fig:top0}
\end{figure}

The production of top quark pairs takes place through quark and gluon
fusion, being the last one the dominating process at the LHC energy
due to the large gluon-gluon luminosity.  We list in Table
\ref{t:top1} the cross sections $\sigma_{\text{bare}}$ and
$\sigma_{\text{pair}}$ evaluated with the same cuts used for the QCD
processes and assuming the top mass to be 174 GeV.  As we can see from
this table, top quark pairs will be copiously produced at LHC, being
the top quark production one of the largest backgrounds before
applying further cuts. Moreover, the $e^\pm$ produced in the top decay
into $be\nu_e$ channel are rather isolated, as shown in Fig.\ 
\ref{fig:top0}.

\begin{figure}
\begin{center}
\mbox{\epsfig{file=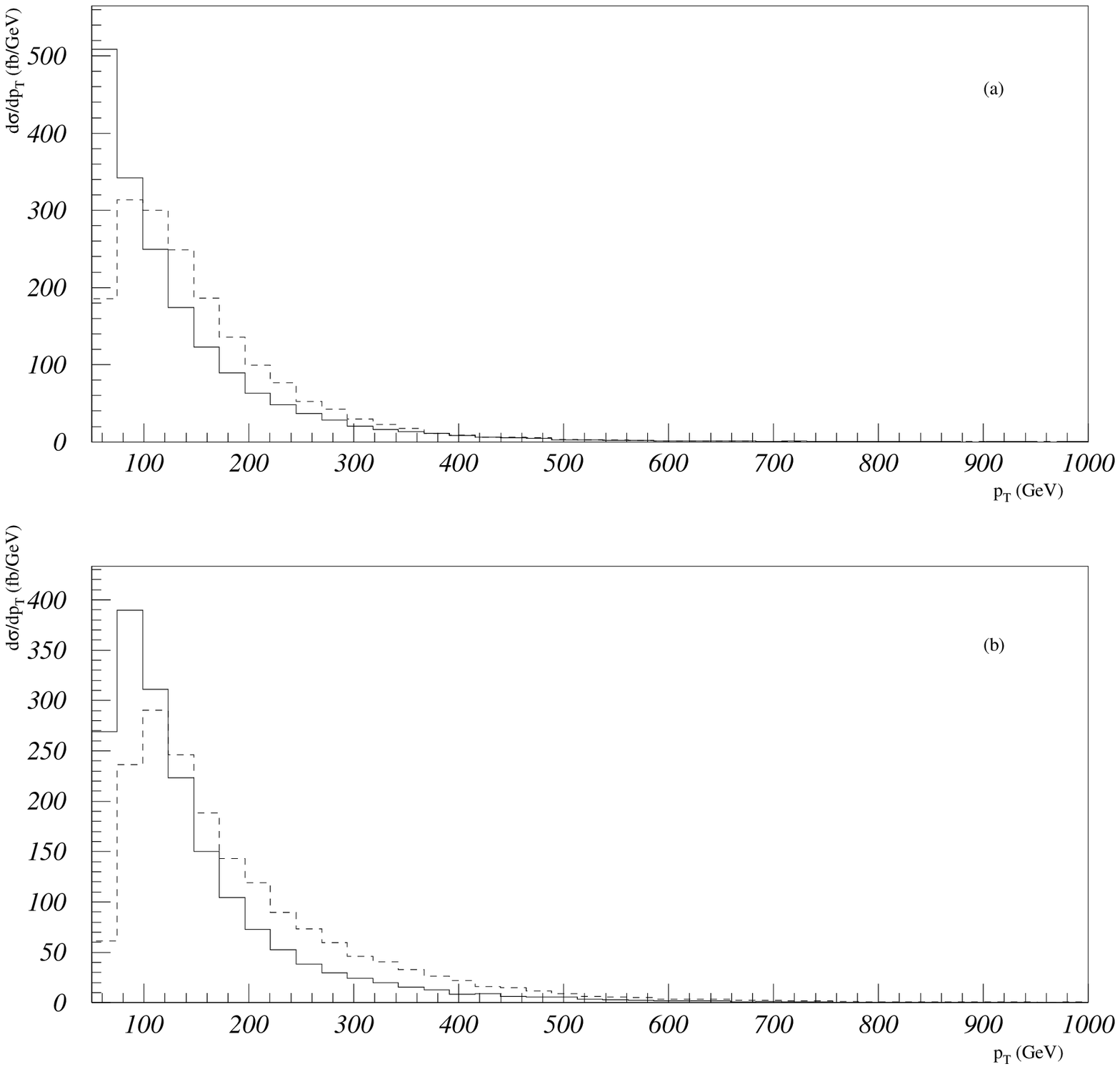,width=.85\linewidth}}
\end{center}
\caption{
  The same distribution of Fig.\ \protect\ref{fig:qcd1} for top
  production. }
\label{fig:top1}
\end{figure}

\begin{figure}
\begin{center}
\mbox{\epsfig{file=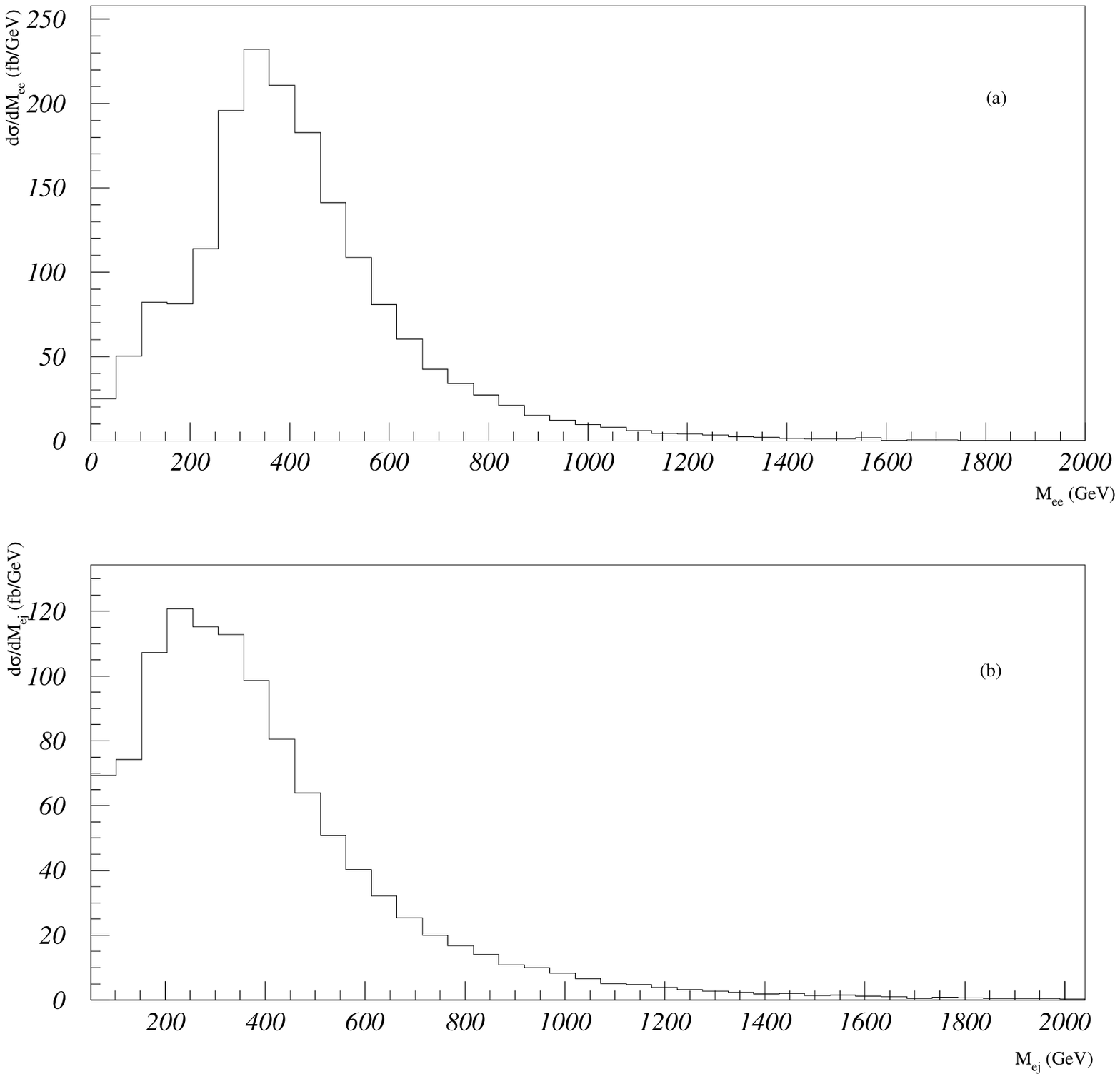,width=.85\linewidth}}
\end{center}
\caption{
  The same distribution of Fig.\ \protect\ref{fig:qcd2} for top
  production. }
\label{fig:top2}
\end{figure}

We present the $p_T$ spectrum of the jets and $e^\pm$ in Fig.\ 
\ref{fig:top1}. As we can see, this distribution is peaked towards low
transverse momenta and contains high $p_T$ leptons (jets) up to 500
(900) GeV. The invariant mass of the $e_1 e_2$ pairs generated in top
decays reaches a maximum around 400 GeV and extends up to 1300 GeV, as
we can see from Fig.\ \ref{fig:top2}a.  From the distribution of
$e_i$--$j_k$ invariant masses, shown in Fig.\ \ref{fig:top2}b, we can
see that this processes can give rise to background events even for
leptoquarks with masses in the TeV range.
 

\section{Signals for the Production of Leptoquarks}
\label{sign}

In hadronic colliders, leptoquarks can be single or pair produced
through the processes
\begin{eqnarray}
q + g &&\rightarrow S_{\text{lq}} + \ell \;\; ,
\label{eq:sin}
\\
q + \bar{q} &&\rightarrow S_{\text{lq}} + \bar{S}_{\text{lq}} \;\; ,
\label{eq:qq}
\\
g + g &&\rightarrow S_{\text{lq}} + \bar{S}_{\text{lq}} \;\; ,
\label{eq:gg}
\end{eqnarray}
where $\ell = e^\pm$ ($\nu$) and we denoted the scalar leptoquark by
$S_{\text{lq}}$. In this work we focus our attention on the case that
$\ell$ is a charged lepton. When the leptoquark decays into a
$e^\pm$--$q$ pair, the final state presents a pair $e^+e^-$ and jets
after hadronization. The cross sections for pair production are model
independent because the leptoquark--gluon interaction is determined by
the $SU(3)_C$ gauge invariance. On the other hand, the single
production is model dependent once it involves the the unknown Yukawa
coupling of leptoquarks to a lepton--quark pairs.


\subsection{Single production}

The cross section for the associated production of a leptoquark and a
charged lepton, process (\ref{eq:sin}), depends quadratically on the
parameter $\kappa$ defined in Eq.\ (\ref{eq:kap}). We exhibit in Table
\ref{t:sin1} the total cross section for the production of a
leptoquark ($\sigma_{\text{pair}}$) that couples only to a charged
lepton and a quark, assuming $\kappa=1$ and requiring that the $e^\pm$
have transverse momenta larger than 50 GeV. As could be anticipated
from the effective Lagrangian (\ref{e:int}), the cross sections for
the production of $e^+q$ and $e^-q$ leptoquarks are equal.
Furthermore, it is interesting to notice that the cross section for
the production of a leptoquark coupling only to $u$ quarks is
approximately twice the one of leptoquarks coupling only to $d$
quarks, in agreement with a naive valence-quark-counting rule.

\begin{figure}
\begin{center}
\mbox{\epsfig{file=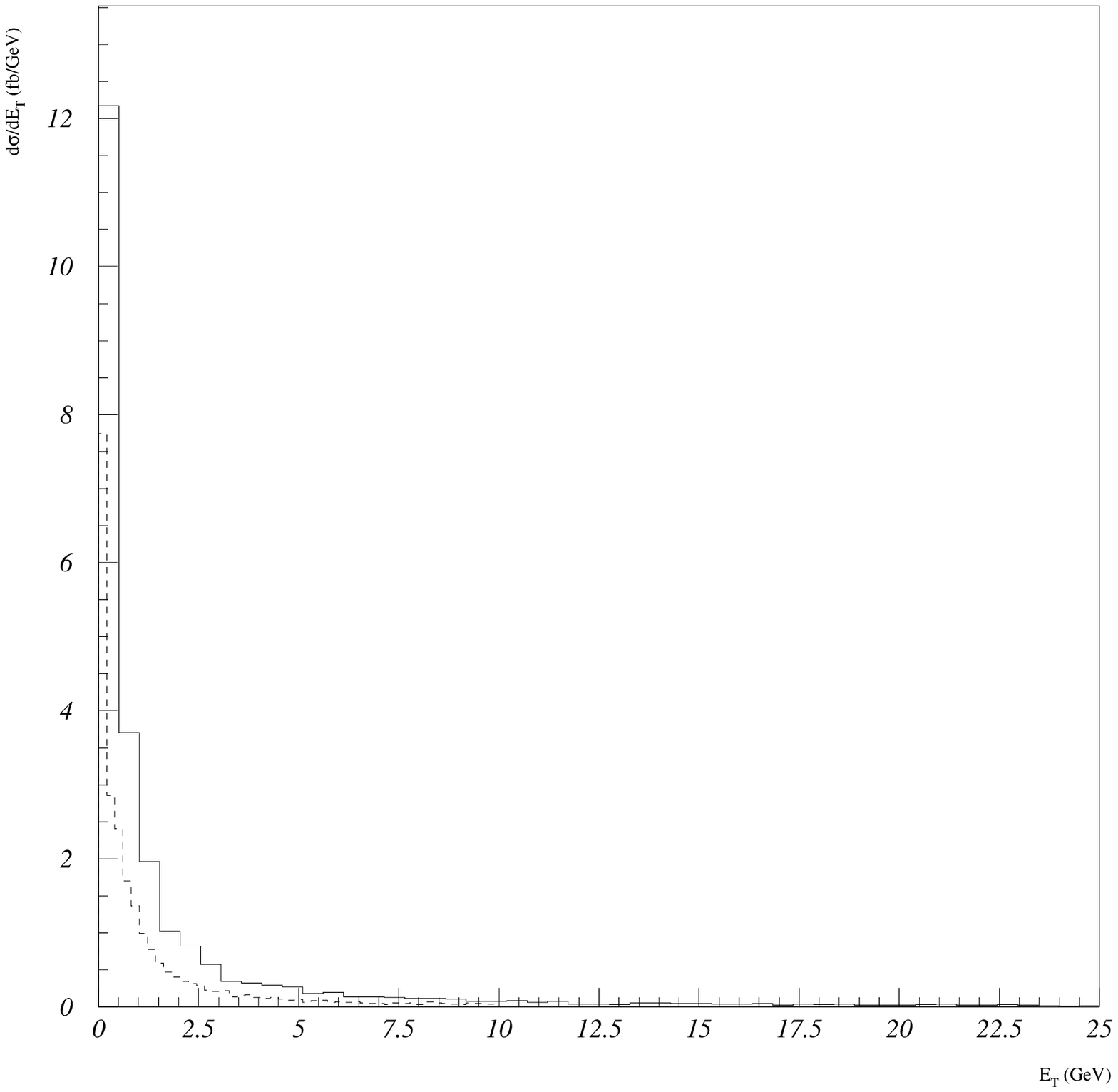,width=.85\linewidth}}
\end{center}
\caption{
  The same distribution of Fig.\ \protect\ref{fig:qcd0} for the single
  production of $e^+\bar{u}$ leptoquarks of mass 1 TeV. }
\label{fig:sing0}
\end{figure}

\begin{figure}
\begin{center}
\mbox{\epsfig{file=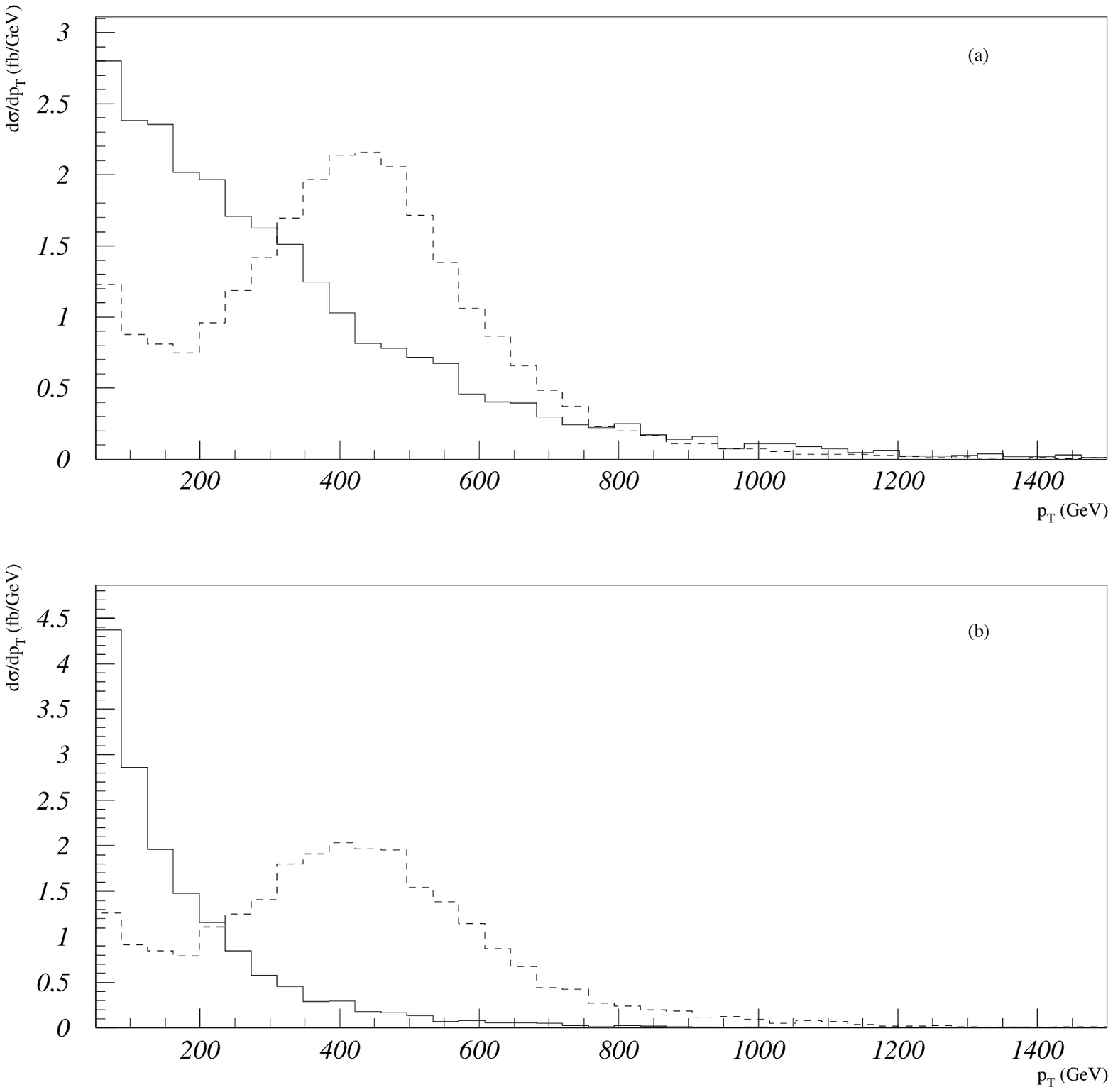,width=.85\linewidth}}
\end{center}
\caption{
  The same distribution of Fig.\ \protect\ref{fig:qcd1} for the single
  production of $e^+\bar{u}$ leptoquarks of mass 1 TeV. }
\label{fig:sing1}
\end{figure}

In what follows, we show several kinematical distributions for
leptoquarks that couple only to $e^+\bar{u}$ pairs and mass $M_{\text
  lq} =1$ TeV, however, the distributions for other leptoquarks are
very similar.  As expected, the bulk of the $e^\pm$ originating from
the single (and also pair) production of leptoquarks are isolated, as
we can see from Fig.\ \ref{fig:sing0}.  We plot in Fig.\ 
\ref{fig:sing1} the $p_T$ spectrum of the leptons $e_{1(2)}$ and jets
$j_{1(2)}$. Notice that the $p_T$ spectrums of the most energetic jet
and $e^\pm$ peak around $M_{\text lq}/2$~($\simeq ~500$ GeV), which
indicates that the $e^\pm$ and jet resulting from the leptoquark decay
are often the ones with the largest transverse momentum in the case of
single leptoquark production. Contrary to what we have seen in the
background processes, the single leptoquark production leads to
$e^\pm$ with large transverse momenta.

\begin{figure}
\begin{center}
\mbox{\epsfig{file=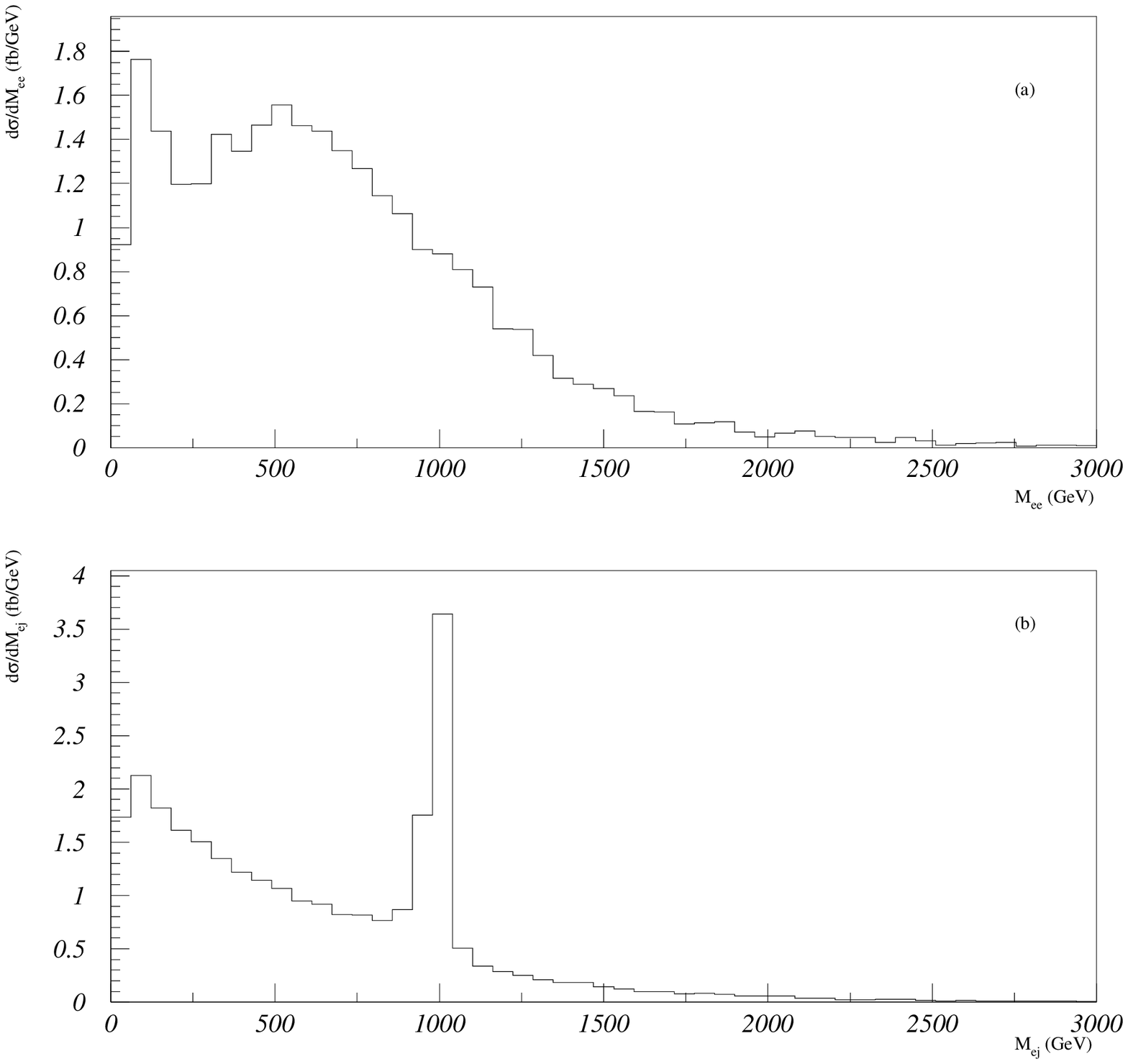,width=.85\linewidth}}
\end{center}
\caption{
  The same distribution of Fig.\ \protect\ref{fig:qcd2} for the single
  production of $e^+\bar{u}$ leptoquarks of mass 1 TeV. }
\label{fig:sing2}
\end{figure}

Fig.\ \ref{fig:sing2}a shows the invariant mass distribution of the
lepton pair $e_1 e_2$, which is peaked toward low invariant masses and
it extends up to very large invariant masses. This feature is
important to reduce the backgrounds, specially the on-shell production
of $Z$'s. Furthermore, the signal for the production of leptoquarks is
a peak in the invariant mass ($M_{ik}$) of $e_i$--$j_k$ pair at
$M_{\text lq}$ (= 1 TeV) which can be easily seen in Fig.\ 
\ref{fig:sing2}b.


\subsection{Pair production}

The pair production of leptoquarks takes place through quark--quark
and gluon--gluon fusions [Eqs.\ (\ref{eq:qq}) and (\ref{eq:gg})],
which rely exclusively on the color gauge interactions of the
leptoquarks, and consequently they do not depend on the leptoquark
species. As we can see from Table \ref{t:pair1}, the gluon--gluon
mechanism dominates the production of leptoquark pairs as long as the
signal is observable at the LHC. Moreover, for $\kappa=1$, the cross
section for the production of leptoquark pairs is larger than the
single production one for light leptoquarks ($M_{\text lq} \lesssim
500$ GeV) due to the large gluon--gluon luminosity.

\begin{figure}
\begin{center}
\mbox{\epsfig{file=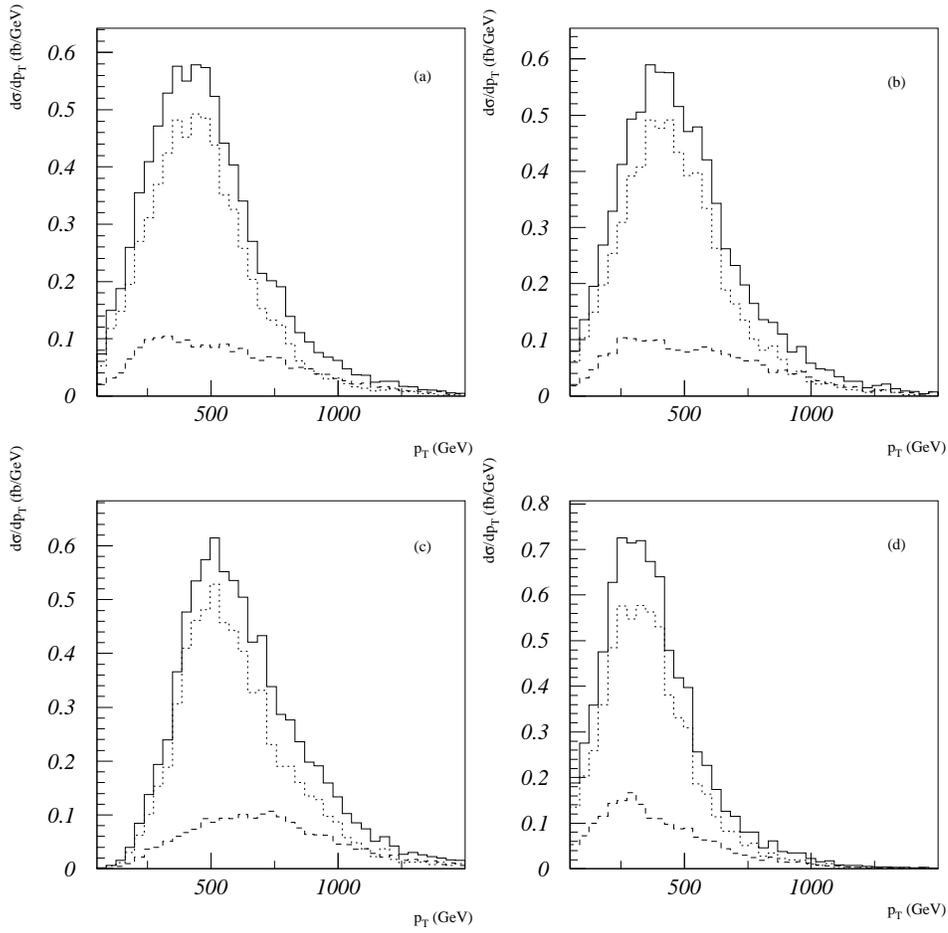,width=.85\linewidth}}
\end{center}
\caption{
  $p_T$ distribution of (a) $e_1$; (b) $e_2$; (c) $j_1$; (d)
  $j_2$; in the pair production of $e^+\bar{u}$ leptoquarks of mass 1
  TeV. The dashed (dotted) line stands for the $q\bar{q}$- ($gg$-)
  fusion contribution while the solid line represents the total
  distribution.  }
\label{fig:pair1}
\end{figure}

The transverse momentum distributions of the $e_{1(2)}$ and $j_{1(2)}$
are shown in Fig.\ \ref{fig:pair1}, where we just required the $e^\pm$
to have $p_T > 50$ GeV. We can see from this figure that the
$q\bar{q}$ and $gg$ processes lead to a similar spectrum that is
peaked approximately at $M_{\text lq}/2$ ($=500$ GeV) and also
exhibits a large fraction of very hard jets and leptons. The presence
of this peak indicates that the two hardest jets and leptons originate
from the decay of the leptoquark pair. However, we still have to
determine which are the lepton and jet coming from the decay of one of
the leptoquarks.

\begin{figure}
\begin{center}
\mbox{\epsfig{file=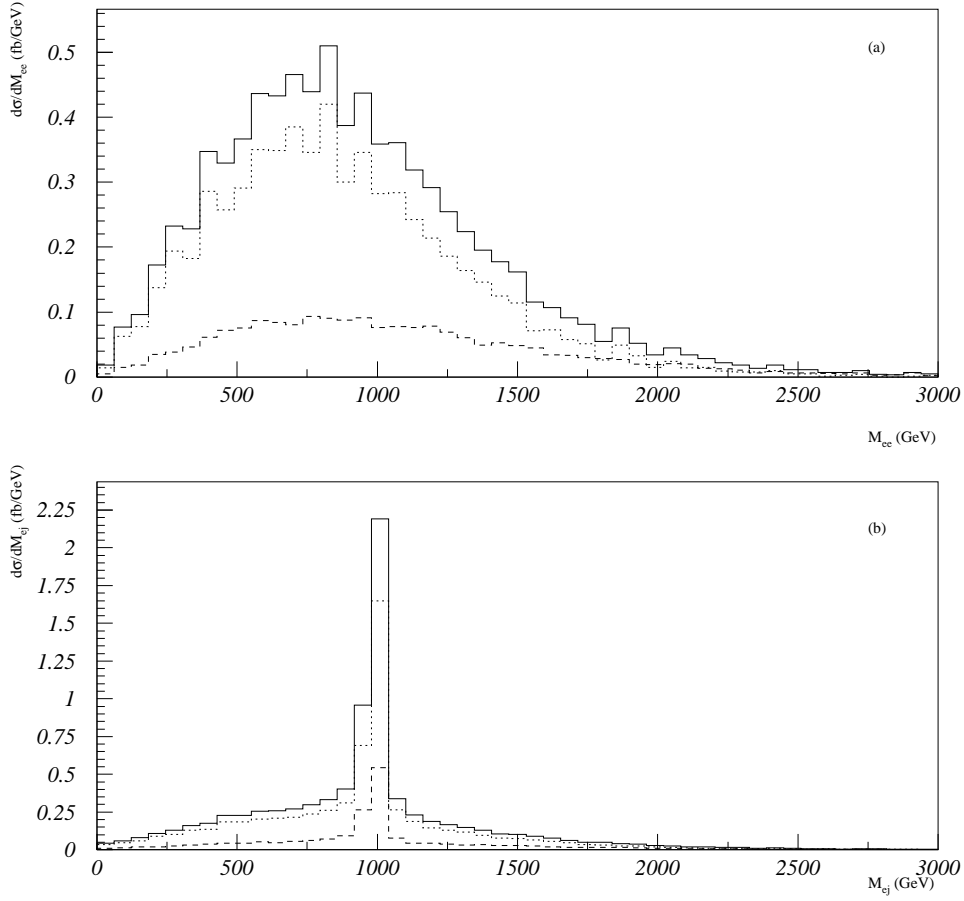,width=.85\linewidth}}
\end{center}
\caption{
  (a) $e^+e^-$ invariant mass distribution; (b) $e^\pm$-jet invariant
  mass spectrum adding the 4 possible combinations for pair production
  of $e^+\bar{u}$ leptoquarks with mass $M_{\text lq} = 1$ TeV. We use the
  conventions of Fig.\ \protect\ref{fig:pair1}.  }
\label{fig:pair2}
\end{figure}

The invariant mass of pairs $e^+ e^-$ is usually quite large as shown
in Fig.\ \ref{fig:pair2}a, and we verified that the invariant mass
spectrum of the $j_1 j_2$ pairs is similar to the $e^+ e^-$ one. There
is a clear peak in the invariant mass ($M_{ik}$) distribution of
$e_i$--$j_k$ pairs, as shown in Fig.\ \ref{fig:pair2}b.


\section{Results}
\label{resu}

There are several goals in our analyses. First, we study the LHC
potentiality to discover a leptoquark, which requires establishing the
cuts to enhance the signal and reduce the backgrounds. Then, we
separate the single and pair productions of leptoquarks in order to be
able to analyze their Yukawa couplings.

Taking into account the features of the signal and backgrounds
described in the previous sections, we imposed the following set of
cuts:

\begin{itemize}
  
\item [(C1)] The first requirement is that the leading jets and
  $e^\pm$ are in the pseudorapidity interval $|\eta| < 3$.
        
\item [(C2)] The leading leptons ($e_1$ and $e_2$) should have $p_T >
  200$ GeV.
      
\item [(C3)] We reject events where the invariant mass of the pair
  $e^+e^-$ ($M_{e_1 e_2}$) is smaller than 190 GeV. This cut reduces
  the backgrounds coming from $Z$ decays into a pair $e^+ e^-$.
      
\item [(C4)] In order to further reduce the $t\bar{t}$ and remaining
  off-shell $Z$ backgrounds \cite{note}, we required that {\em all}
  the invariant masses $M_{e_i j_k}$ are larger than 200 GeV, since
  pairs $e_i j_k$ coming from an on-shell top decay have invariant
  masses smaller than $m_{\text top}$. The present experiments are
  able to search for leptoquarks with masses smaller than 200 GeV,
  therefore, this cut is in agreement with the present bounds.

\end{itemize}

In principle we should also require the $e^\pm$ to be isolated from
hadronic activity in order to reduce the QCD backgrounds.
Nevertheless, we verified that our results do not change when we
introduced typical isolation cuts in addition to any of the above
cuts. In fact, the cuts C1 and C2 eliminate all the QCD processes,
except for $gg \rightarrow q \bar{q}$, for which 0.25\% of the events
survive the cuts C1 and C2 (and also C3). However, this background
disappears when we apply C4.

We exhibit in Table~\ref{t:ew1} the fraction of electroweak events
that passes the cuts C1, C2, and C3.  As we can see from this table,
the electroweak backgrounds associated to the production of a $Z$ or a
$W^+W^-$ pair survive the cuts C1 and C2 and they are not completely
eliminated even when we apply the cut C3. At this point, the most
important backgrounds in this class are the production of a $Z$ in
association with a quark or a gluon, which have cross sections of the
order of a few fb. Analogously to the QCD events, the electroweak
backgrounds disappear after imposing C4.

A substantial fraction of the top backgrounds passes the cuts C1 and
C2, as we can see from Table~\ref{t:top1}. The $e^+e^-$ invariant mass
cut (C3) reduces these backgrounds, however, they remain rather large,
being of the order of 50 fb. When the C4 cut is applied, the only
events that pass all cuts stem from $gg \rightarrow t \bar{t}$;
nevertheless, the cross section of this process is smaller than
$10^{-3}$ fb when we integrate in an invariant mass bin around a
leptoquark mass; see below.

Since the cuts C1---C4 effectively suppress all the backgrounds to
leptoquark production at the LHC, the leptoquark searches are free of
backgrounds. Therefore, the LHC will be able to exclude with 95\% CL
the regions of parameter space where the number of expected signal
events is larger than 3 for a given integrated luminosity.


\subsection{Leptoquark pair search}

In our analysis of leptoquark pair production we applied the cuts
C1---C4 and also demanded the events to exhibit two $e$-jet pairs with
a invariant masses in the range $| M_{lq} \pm \Delta M|$ with $\Delta
M$ given in Table~\ref{bins}. The limits on the leptoquarks depend
exclusively on its branching ratio into a charged lepton and jet
($\beta$) since they are produced by strong interaction processes that
are the same for all leptoquark species. We show in Table~
\ref{t:lim:pair} the 95\% CL limits on the leptoquark masses that can
be obtained from their pair production at the LHC for two different
integrated luminosities. As we can see, this search will be able to
exclude leptoquarks with masses up to 1.5 (1.7) TeV for $\beta = 0.5$
(1) and an integrated luminosity of 100 fb$^{-1}$, increasing
considerably the present bounds.


\subsection{Single leptoquark search}

We also analyzed the search for leptoquarks through the existence of
an excess of events presenting a $e$-jet invariant mass in the range
$| M_{lq} \pm \Delta M|$ after we imposed the cuts C1---C4; see
Table~\ref{bins} for $\Delta M$. These events originate from the
single production of leptoquarks, as well as their pair production,
consequently having a larger cross section than the pair or single
production alone, and then increasing the reach of the LHC. However,
this search is model dependent since the single production involves
the couplings of the leptoquark to fermions.  In fact, the cross
section is determined by $\kappa$, $\beta$, and the quark that the
leptoquark couples to. Therefore, for fixed $\kappa$, the production
cross sections for leptoquarks $S_{1L}$ and $S_3^0$ [($S_{1R}$,
$R^1_{2L}$, and $R^1_{2R}$) or ($S^+_3$, $R^2_{2R}$, $\tilde{R}^1_2$,
and $\tilde{S}_{1R}$) ] are equal.

\begin{figure}
\begin{center}
  \mbox{\epsfig{file=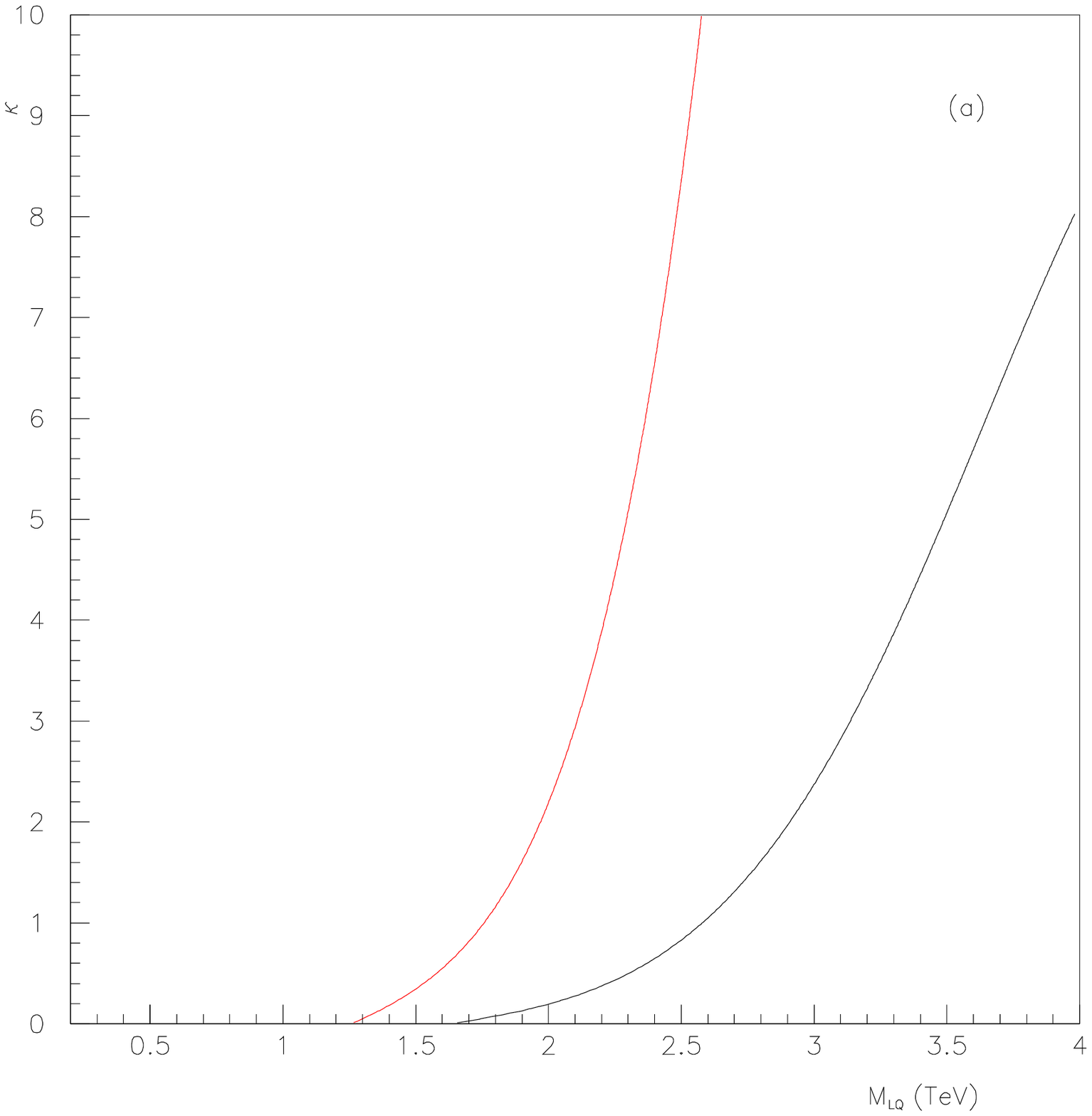,width=0.4\linewidth}}
\end{center}
\begin{center}
  \mbox{\epsfig{file=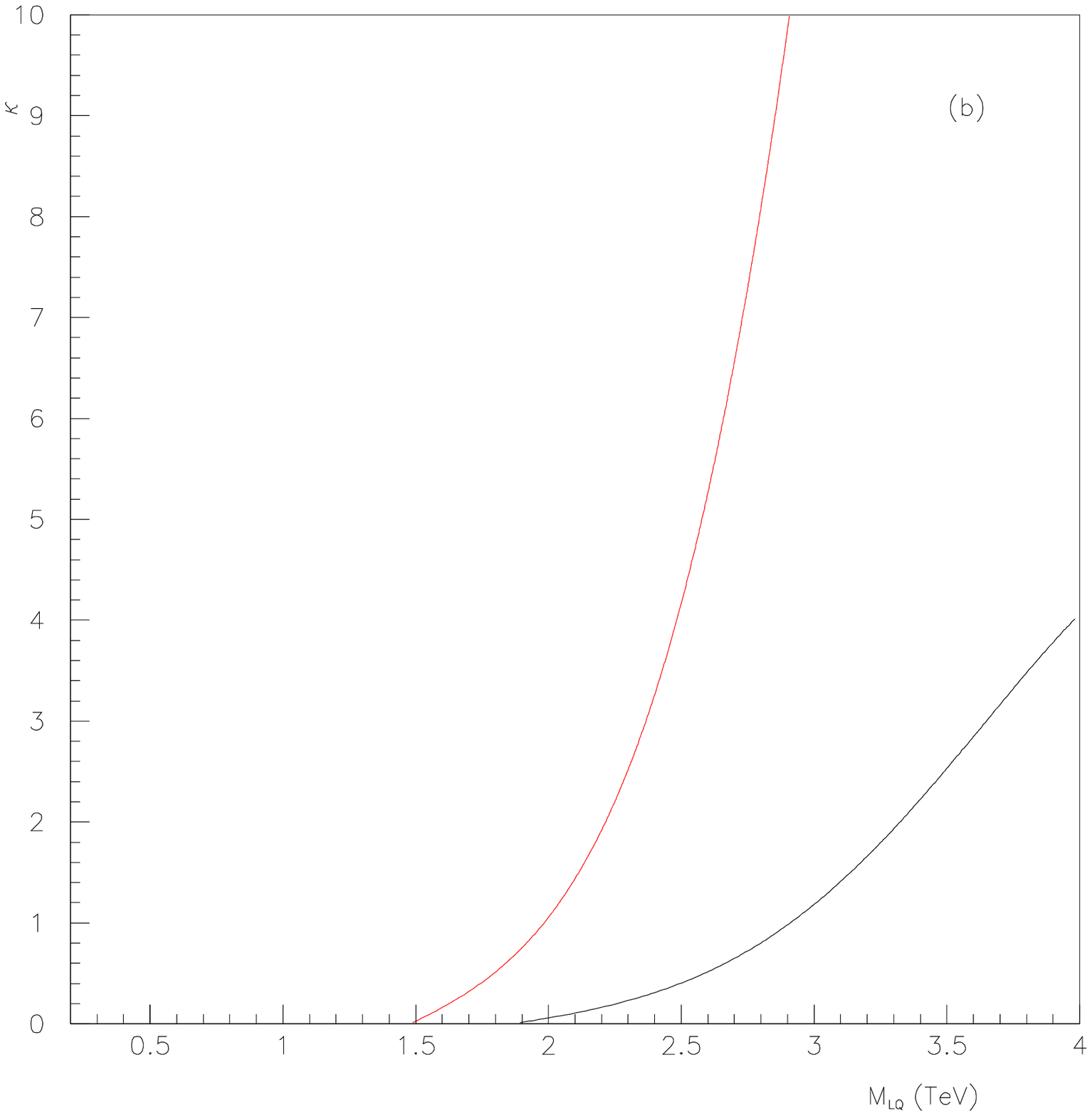,width=0.4\linewidth}}
\end{center}
\begin{center}
  \mbox{\epsfig{file=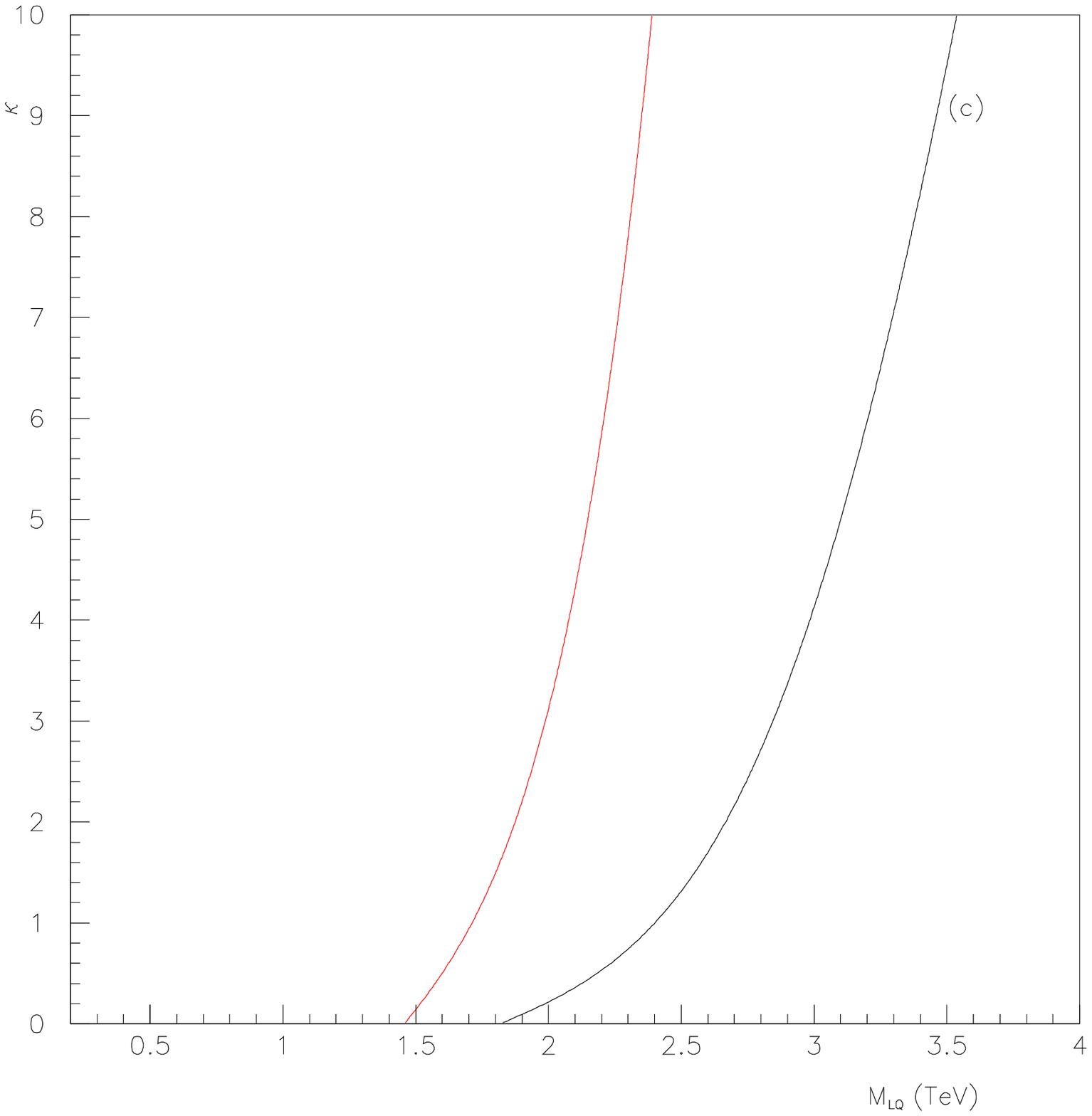,width=0.4\linewidth}}
\end{center}
\caption{
  95\% excluded regions in the plane $\kappa$--$M_{lq}$ from the
  single leptoquark analysis for an integrated luminosity of 10/100
  fb$^{-1}$ (solid/dotted line) and the leptoquarks: (a) $S_{1L}$ and
  $S_3^0$; (b) $S_{1R}$, $R^1_{2L}$, and $R^1_{2R}$; (c)$S^+_3$,
  $R^2_{2R}$, $\tilde{R}^1_2$, and $\tilde{S}_{1R}$.  }
\label{kap_mlq}
\end{figure}

Fig.\ \ref{kap_mlq} contains the 95\% CL excluded regions in the plane
$\kappa$--$M_{lq}$ from the single leptoquark search.  As expected,
the excluded region is independent of $\kappa$ for masses up to the
reach of the leptoquark pair searches; see Table~ \ref{t:lim:pair}. At
higher masses the signal is dominated by the single production and
consequently the bounds on leptoquarks that couple to $d$ quarks
($S^+_3$, $R^2_{2R}$, $\tilde{R}^1_2$, and $\tilde{S}_{1R}$) are the
weakest ones for a fixed value of $\kappa$. Since the leptoquarks
$S_{1R}$, $R^1_{2L}$, and $R^1_{2R}$ couple to $u$ quarks and have
$\beta=1$ they are the ones that possesses the most stringent limits.
In fact, for leptoquark Yukawa couplings of the electromagnetic
strength ($\kappa=1$) and an integrated luminosity of 100 fb$^{-1}$,
the LHC can exclude $S_{1L}$ and $S_3^0$ leptoquarks with masses
smaller than 2.6 TeV; while $S^+_3$, $R^2_{2R}$, $\tilde{R}^1_2$, and
$\tilde{S}_{1R}$ leptoquarks with masses smaller than 2.4 TeV can be
ruled out; and $S_{1R}$, $R^1_{2L}$, and $R^1_{2R}$ leptoquarks can be
excluded up to masses of 2.9 TeV.


\section{Conclusions}
\label{conc}

The discover of leptoquarks is without any doubt a striking signal for
the existence of life beyond the standard model. The LHC will be able
to discover leptoquarks with masses smaller than 1.5--1.7 TeV
irrespective of their Yukawa couplings through their pair production
for an integrated luminosity of 100 fb$^{-1}$. Furthermore, the single
leptoquark search can extend the reach of the LHC, allowing the
discovery of leptoquarks with masses up to 2--3 TeV depending on their
couplings to fermions. It is interesting to notice that the single
leptoquark searches not only enlarge the reach of a collider like the
LHC, but also allows to study the Yukawa couplings of the leptoquarks
in a greater detail.  For instance, from the numbers of events with
$e^+$-jet and $e^-$-jet we can infer whether the leptoquark couples to
$e^-\bar{q}$ or $e^+\bar{q}$. From the size of the single production
cross section we can also estimate $\kappa$ for leptoquarks coupling
to $u$ and $d$ quarks.  Therefore, the LHC will be able to either make
a detailed study of leptoquarks or improve considerably the present
bounds.

\acknowledgments

We would like to thank prof.\ T.\ Sj\"ostrand for valuable comments
and discussions.  One of us (O.J.P.E.) would like to thank the kind
hospitality of the Institute for Elementary Particle Research,
University of Wisconsin--Madison, where the part of this work was
done. This work was partially supported by Conselho Nacional de
Desenvolvimento Cient\'\i fico e Tecnol\'ogico (CNPq), and
Funda\c{c}\~ao de Amparo \`a Pesquisa do Estado de S\~ao Paulo
(FAPESP).



\begin{table}[htbp]
\begin{center}
\begin{tabular}{|c|c|c|c|}
leptoquark & decay & branching ratio & $\kappa 4\pi\alpha_{\text{em}}$ 
\\
\hline
$S_{1L} $         & $e^+ \bar{u}$ &  50\%  & $\frac{g_{1L}^2}{2}$ \\
$S_{1R} $         & $e^+ \bar{u}$ & 100\%  & $\frac{g_{1R}^2}{2}$ \\
$\tilde{S}_{1R}$  & $e^+ \bar{d}$ & 100\%  & $\frac{\tilde{g}_{1R}^2}{2}$ \\
$S^+_3$           & $e^+ \bar{d}$ & 100\%  & $g_3^2$ \\
$S^0_3$           & $e^+ \bar{u}$ &  50\%  & $\frac{g_3^2}{2}$   \\
$R_{2L}^1$        & $e^- \bar{u}$ & 100\%  & $\frac{h_{2L}^2}{2}$ \\
$R_{2R}^1$        & $e^- \bar{u}$ & 100\%  & $\frac{h_{2R}^2}{2}$ \\
$R_{2R}^2$        & $e^- \bar{d}$ & 100\%  & $\frac{h_{2R}^2}{2}$ \\
$\tilde{R}_2^1$   & $e^- \bar{d}$ & 100\%  & $\frac{\tilde{h}_{2L}^2}{2}$ \\
\end{tabular}
\vskip 0.75cm
\caption{Leptoquarks that can be observed through their decays into
  a $e^\pm$ and a jet and the correspondent branching ratios into this
  channel. We also show the relation between the leptoquark Yukawa
  coupling and the parameter $\kappa$ used in PYTHIA.}
\label{t:cor}
\end{center}
\end{table}


\begin{table}[htbp]
\begin{center}
\begin{tabular} {|c|c|c|}
Process  &  $\sigma_{\text{bare}}$ (nb)   & $\sigma_{\text{pair}}$ (fb) 
\\
\hline
$q_i q_j~ \to q_i q_j$ & 1.0~10$^{2}$ & 4.8~10$^2$
\\
$q_i \bar{q}_i~ \to q_k \bar{q}_k$ & 1.8 & 1.2~10$^2$
\\
$q_i \bar{q}_i~ \to g g$  & 1.6 & 2.3
\\
$q_i g~ \to q_i g $ & 6.2~10$^2$ & 1.0~10$^3$
\\
$g g~ \to q_k \bar{q}_k $ & 25. & 1.2~10$^3$
\\
$g g~ \to g g $ & 6.9~10$^2$ & 5.5~10$^2$
\\
\end{tabular}
\vskip 0.75cm
\caption{Background processes included in the QCD class and their
  respective cross sections. To obtain $\sigma_{\text bare}$ we
  required that the hard scattering process has $p_T > 100$ GeV; for
  $\sigma_{\text pair}$ we further demanded that the $e^\pm$ have $p_T
  > 50$ GeV.}
\label{t:qcd1}
\end{center}
\end{table}

\begin{table}[htbp]
\begin{center}
\begin{tabular} {|c|c|c|c|c|}
Process  &  $\sigma_{\text{bare}}$ (pb)  & $\sigma_{\text{pair}}$ (fb)
& $\epsilon_{12}$  & $\epsilon_{123}$
\\
\hline
$ q_i \bar{q}_i~ \to g \gamma $      & 74.         & 0.14
& 0     &  0
\\
$ q_i \bar{q}_i~ \to g Z  $          & 95.         & 9.2~10$^2$
& 1\%   &  0.8\%
\\
$ q_i \bar{q}_j~ \to g W^\pm $       & 2.2~10$^2$  & 8.8 
& 0   & 0
\\
$ q_i \bar{q}_i~ \to \gamma Z $      & 1.4         & 16.
& 0.6\%  & 0.4\%
\\
$ q_i \bar{q}_j~ \to \gamma W^\pm $  & 1.1         & 6.6~10$^{-3}$
& 0  & 0
\\
$ q_i \bar{q}_i~ \to Z Z $           & 1.4         & 29.
& 0.8\%  & 0.2\%
\\
$ q_i \bar{q}_j~ \to Z W^\pm  $      & 2.9         & 37.
& 1.2\% & 0.4\%
\\
$ q_i \bar{q}_i~ \to W^+ W^- $       & 6.8         & 38.
& 4.3\%  & 4.3\%
\\
$ q_i g~ \to q_i \gamma $            & 6.1~10$^2$  & 3.5
& 0   & 0
\\
$ q_i g~ \to q_i Z $                 & 5.5~10$^2$  & 5.4~10$^3$
& 0.8\%  & 0.2\%
\\
$ q_i g~ \to q_k W^\pm   $           & 1.4~10$^3$  & 2.8~10$^2$
& 0.25\%   & 0.25\%
\\
\end{tabular}
\vskip 0.75cm
\caption{Background processes included in the electroweak class and their
  respective cross sections with the same cuts used in
  Table~\protect\ref{t:qcd1}. We also exhibit the fraction of events
  $\epsilon_{12}$ ($\epsilon_{123}$) that survive the cuts C1 and C2
  (C1, C2, and C3); see text.}
\label{t:ew1}
\end{center}
\end{table}


\begin{table}[htbp]
\begin{center}
\begin{tabular} {|c|c|c|c|c|}
Process  &  $\sigma_{\text{bare}}$ (pb)  & $\sigma_{\text{pair}}$ (fb)
& $\epsilon_{12}$  & $\epsilon_{123}$
\\
\hline
$q_i q_j \to t q_k$   & 40.   &  11.
& 0.85\%  & 0.01\%
\\
$q_i \bar{q}_i~ \to t \bar{t}$   & 0.40         &  2.0~10$^2$
& 0.40\%   & 0.39\%
\\
$g g~ \to t \bar{t} $            & 2.9~10$^2$   &  1.4~10$^3$
& 75\%  & 3.5\%
\\
\end{tabular}
\vskip 0.75cm
\caption{Background processes due to top quark production and their
  respective cross sections with the same cuts used in
  Table~\protect\ref{t:qcd1}.  We also exhibit the fraction of events
  $\epsilon_{12}$ ($\epsilon_{123}$) that survive the cuts C1 and C2
  (C1, C2, and C3).}
\label{t:top1}
\end{center}
\end{table}

\begin{table}[htbp]
\begin{center}
\begin{tabular} {|c|c|c|c|c|c|}
$\ell q$ coupling  &  $M_{\text lq}=500$ GeV & 1000 GeV & 1500 GeV & 2000 GeV
& 2500 GeV
\\
\hline
$e^\pm u$ & 5.1~10$^2$ & 27. & 4.1 & 0.98 & 0.29
\\
$e^\pm d$ & 2.9~10$^2$ & 14. & 2.2 & 0.55 & 0.18
\\
\end{tabular}
\vskip 0.75cm
\caption{Total cross section in fb for the single production of a leptoquark
that couples only to pair $\ell q$ for several leptoquark masses. We
required that the produced $e^\pm$ have $p_T > 50$ GeV, 
and that the scattering process has $p_T > 100$ GeV. }
\label{t:sin1}
\end{center}
\end{table}

\begin{table}[htbp]
\begin{center}
\begin{tabular} {|c|c|c|c|c|}
process    &  $M_{\text lq}=500$ GeV & 1000 GeV & 1500 GeV & 2000 GeV 
\\
\hline
$q\bar{q}$ fusion & 86.        & 1.9 & 0.10  & 2.7 10$^{-3}$
\\
$gg$       fusion & 4.9 10$^2$ & 6.3 & 0.25  & 1.5 10$^{-2}$
\\
\end{tabular}
\vskip 0.75cm
\caption{Total cross section in fb for the pair production of  leptoquarks,
  requiring that the produced $e^\pm$ have $p_T > 50$ GeV, and that
  the hard scattering process has $p_T > 100$ GeV. }
\label{t:pair1}
\end{center}
\end{table}


\begin{table}[htbp]
\begin{center}
\begin{tabular} {|c|c|}
$M_{lq}$ (GeV) & $\Delta M$ (GeV)
\\
\hline
 300   &  50 \\
 500   &  50 \\
1000   & 150 \\
1500   & 200 \\
2000   & 200 \\
2500   & 300 \\
\end{tabular}
\vskip 0.75cm
\caption{Invariant mass bins used in our analyses as a function of
  the leptoquark mass.}
\label{bins}
\end{center}
\end{table}


\begin{table}[htbp]
\begin{center}
\begin{tabular} {|c|c|c|}
leptoquark & ${\cal L}=10$ fb$^{-1}$ &${\cal L}=100$ fb$^{-1}$  
\\
\hline
$S_{1L} $ and $S^0_3$   & 1.1 & 1.5 \\
$S_{1R} $,   $\tilde{S}_{1R}$, $R_{2L}^1$,$R_{2R}^2$, and
$\tilde{R}_2^1$  & 1.3 & 1.7 
\end{tabular}
\vskip 0.75cm
\caption{
  95\% CL limits on the leptoquark masses in TeV that can be obtained
  from the search for leptoquark pairs for two integrated
  luminosities.  }
\label{t:lim:pair}
\end{center}
\end{table}



\begin{references}
\bibitem{ckin} This cut on the hard scattering partons is not physical,
  however, it increases the efficiency of the Monte Carlo without
  affecting the high $p_T$ distribution of the final leptons and jets.
\bibitem{note} In our analyses we generated a set of off-shell $Z$
  events, with invariant mass larger than 200 GeV, in order to study
  the tail of the $e^+e^-$ invariant mass distribution.
\end{references}
\end{document}